\documentclass[nonacm,sigconf,screen,11pt]{acmart}

\usepackage{algpseudocode}
\usepackage{listings}
\usepackage{tikz}
\usepackage{xcolor}
\usepackage[ruled,vlined,linesnumbered]{algorithm2e}

\definecolor{codegreen}{rgb}{0,0.6,0}
\definecolor{codegray}{rgb}{0.5,0.5,0.5}
\definecolor{codepurple}{rgb}{0.58,0,0.82}
\definecolor{backcolour}{rgb}{0.95,0.95,0.92}

\lstdefinestyle{mystyle}{
    backgroundcolor=\color{backcolour},   
    commentstyle=\color{codegreen},
    keywordstyle=\color{magenta},
    numberstyle=\tiny\color{codegray},
    stringstyle=\color{codepurple},
    basicstyle=\ttfamily\tiny,
    breakatwhitespace=false,         
    breaklines=true,                 
    captionpos=b,                    
    keepspaces=true,                 
    numbers=left,                    
    numbersep=5pt,                  
    showspaces=false,                
    showstringspaces=false,
    showtabs=false,                  
    tabsize=2,
    frame=ltb,
    framerule=0pt,
}

\lstdefinelanguage{AmbientTalk}{
    morekeywords={def, isolate, init, self, new},
    sensitive=true, 
    morecomment=[l]\#, 
    morestring=[b]",
    morestring=[b]'
}

\lstdefinelanguage{Lua}{
    morekeywords={local, require, function, self},
    sensitive=true, 
    morecomment=[l]\#, 
    morestring=[b]",
    morestring=[b]'
}

\AtBeginDocument{
  }

\setcopyright{acmlicensed}
\copyrightyear{2018}
\acmYear{2018}
\acmDOI{XXXXXXX.XXXXXXX}

\acmConference[Conference acronym 'XX]{Make sure to enter the correct
  conference title from your rights confirmation emai}{June 03--05,
  2018}{Woodstock, NY}

\acmISBN{978-1-4503-XXXX-X/18/06}

\usepackage{xcolor}
\usepackage{amssymb}

\begin{document}

\title{Dynamic Software Updates using CRDTs}

\author{Seppe Wyns}
\affiliation{
  \institution{Vrije Universiteit Brussel}
  \city{Brussels}
  \country{Belgium}
  }
\email{seppe.wyns@vub.be}

\author{Jim Bauwens}
\affiliation{
  \institution{Vrije Universiteit Brussel}
  \city{Brussels}
  \country{Belgium}
  }
\email{jim.bauwens@vub.be}

\author{Elisa {Gonzalez Boix}}
\affiliation{
  \institution{Vrije Universiteit Brussel}
  \city{Brussels}
  \country{Belgium}
  }
\email{egonzale@vub.be}

\renewcommand{\shortauthors}{Seppe Wyns}

\begin{abstract}

This paper investigates how Conflict-free Replicated Data Types (CRDTs) can be used for dynamic software updates of distributed applications. 
We propose to model application updates as a new \emph{App CRDT} that stores the application code associated with a semantic version, which defines a total order of the code updates.  

The App CRDT works with an \emph{API-compatible} message delivery middleware, which allows applications to continue working with partially updated components in the face of backwards-incompatible software updates. 

We implemented our approach in \textit{AmbientTalk}, an ambient-oriented programming language designed for distributed systems. 

We show how this CRDT can be integrated with existing AmbientTalk applications, requiring minimal changes. 
We also implemented our approach in \textit{LuAT}, an ambient-oriented programming framework for Lua. This shows that our approach of using CRDTs to replicate code can be generalised to other programming languages.
\end{abstract}

\maketitle

\section{Introduction} \label{sec:introduction}
Many distributed systems are expected to run continuously, e.g. cloud services, micro-services, large-scale data processing systems, IoT applications like traffic cameras systems, etc. 
Yet, they must often be updated to fix bugs or security vulnerabilities, upgrade their functionality, or apply performance optimisations.

Since many distributed systems use databases to store their data, many migration techniques have been introduced for these systems. These techniques allow developers to move data between servers \cite{prorea, cut-me-some-slack}, update the schema of the data \cite{bullfrog}, and distribute and migrate data sharded across multiple nodes \cite{remus}. 

Migrating data is not the only challenging aspect of updating a distributed system. 
Developers must also ensure that code updates are distributed to all application instances and that communication between different instances running different versions does not break the system. 

In this work, we explore a dynamic software update (DSU) solution \cite{Hicks2005,Pina2013,Tesone2018} for distributed applications. 
DSU manages the migration of a piece of software from one version to another while the application is running without the need for a shutdown and restart. 
Instead, a DSU stops the application at a safe point, installs the new version and restarts it. 
However, DSU is problematic in a distributed setting as it requires locking all the application processes running on different devices at specified synchronization points to apply code changes. 
This is costly and affects the overall availability of the system.

In this paper, we propose DSU approach built on the work on Replicated Data Types (RDTs)~\cite{Shapiro2011,Burckhardt12,Baquero17,conflict-free-replicated-datatypes}, 
to allow software updates to be propagated asynchronously to the different application processes in the distributed system. 
RDTs offer a principle approach to guarantee convergence without requiring synchronisation which can be applied to guarantee that all processes in a distributed system are updated. 
In particular, we propose handling DSU using a \textit{conflict-free replicated data type} (CRDT)\cite{Shapiro2011}.

We introduce a novel CRDT for replicating application code called \emph{App CRDT}, which ensures that code updates in a distributed system are eventually applied to all application processes. 
Our DSU approach does not require all updates to be compatible:
the App CRDT works with an \emph{API-compatible message delivery} middleware to ensure that communication between application processes running different code versions do not break the overall distributed system.

Our design also allows developers to specify how the application state should be migrated, ensuring no data is lost during an update.

We have implemented our CRDT-based DSU approach in AmbientTalk\cite{ambienttalk}, an ambient-oriented programming language designed for distributed systems.

We validated our approach by exploring several sample applications built using our AmbientTalk prototype implementation. 

We additionally present a second implementation of our approach in LuAT\cite{luat}, an ambient-oriented programming framework for Lua, highlighting that our approach is not only applicable to purely academic programming languages but also to real-world systems.

\section{Design Concerns for a DSU Approach for Distributed Systems} 
\label{sec:design-goals}
The overall goal is to design a DSU approach capable of updating a distributed application, minimising the downtime of the application during the update and guaranteeing it continues working while the different distributed components are progressively updated. 
The idea of our work is to employ a CRDT at the core of the DSU to propagate and apply code updates to all components in a distributed application without requiring coordination.

This section describes the design goals for such a CRDT-based DSU approach.

\subsection{Application Independence} \label{sec:application-independece}
The DSU approach should apply to any distributed application. 
We will use a CRDT to replicate the application code amongst the different nodes of the distributed application.
Even though the application code is stored within the CRDT, the CRDT design should not impose any restriction on the type of application that can be updated. 

In other words, the CRDT should be independent of the application code it replicates, ensuring compatibility with distributed applications regardless of their implementation.

\subsection{Software Update Transparency} \label{sec:update-transparency}

All communication between the CRDT replicas that propagate software updates should be transparent to the implementation of the application itself. 
The application code should not need to contain any logic that checks whether it is up-to-date, and it should not be concerned with distributing updates across different components of the application. 
The CRDT implementation itself hides the complexity of the update distribution logic. 
However, the DSU approach should also ensure that application-specific messages sent between the different components of the distributed application do not break the overall application execution, even if some components have not been yet updated and run an older version of the code.

\subsection{State Migration} \label{sec:state-migration}

As a CRDT will manage all the application's code updates, an application does not have control over when it might be updated. 
When the application needs to be updated, the DSU should offer a way to migrate the application's state from an old code version to a new one. 

However, the logic of state migration is application-dependent and cannot be inferred from code changes\cite{Tesone2018}. 
The DSU approach should thus provide the means for developers to specify what needs to be migrated.

\subsection{Separation of Concerns} \label{sec:separation-of-concerns}
There should be a clear distinction between the application code and the logic needed to update the application itself. 
The implementation of the application should not be mixed with the logic required to migrate the state, for example. Although migrating the state of an application is inherently intertwined with the application itself, the developer should be able to separate the logic of updating their application from the actual implementation itself.

\section{CRDT-based DSU Approach}
\label{sec:approach}

Before introducing our CRDT-based approach to dynamic software updates of distributed applications, we define our system model. 

We consider a distributed application running on multiple devices, each device running the same code. 
We call an \emph{application instance} to an executing process on a device running the application code. 
An application instance offers several operations and has an associated state. 
The application state comprises a set of objects, which can be accessed and modified through the application operations. 
An operation consists of reads and writes on the local state executed atomically. 
Operations cannot modify the state of objects on other application instances. 
Instead, the operation must send an asynchronous message to the other application instance. 
Upon receiving a message, the system delivers it to the application, which results in the local execution of the corresponding operation for the message. 
We assume applications communicate with one another over eventual-reliable point-to-point channels and a fail-recover model in which instances may crash and later recover. We do not consider Byzantine faults.

In practice, such a system model can be implemented by a distributed actor-based framework, e.g. Akka, or language, e.g. Ambientalk (cf. Section \ref{sec:implementation}). 

\subsection{CRDT Specification} \label{sec:specification}

We model code updates with a state-based CRDT in which the payload is the application code. 

We model the application code as closures, but other representations are possible ( e.g. JavaScript uses strings, Julia uses symbolic expressions, etc.).
We chose a state-based CRDT design as it requires fewer assumptions from the underlying framework or programming language. Modelling code updates with an operation-based CRDT design requires the programming language to reify code changes and dynamically apply them to a running system. This is only possible in some languages with advanced DSU support, such as Smalltalk or CLOS.

To allow the CRDT to differentiate code updates, we associate a version with the application code. 
Versions are assumed to be totally orderable. Our approach uses semantic versions\footnote{\url{https://semver.org}} to represent application versions, although another scheme may be used.

A semantic version consists of three subversions: major, minor, and patch. An increment to the patch version represents fixes to the application that do not provide new functionality. A new minor version indicates the addition of a new feature while still being backwards compatible. Finally, a major version increment represents a backwards-incompatible API change. Semantic versions allow us to define a total order of code updates, enabling the CRDT to decide what updates to apply.

\begin{algorithm}
  \SetKw{State}{state:}{}{}
  \SetKwProg{On}{on}{ :}{}
  \SetKwProg{Query}{query}{}{}
  \SetKwProg{Update}{update}{}{}
  \SetKwProg{Merge}{merge}{}{}
  \DontPrintSemicolon
  \SetKwComment{Comment}{$\triangleright$ }{}
  \SetAlgoLined
  \State {   \\
     \quad $V := 0.0.0$ \Comment*[r]{V : semantic version} 
     \quad $C := nil$ \Comment*[r]{C : Code} 
  } 
  \Query {getValue() : (Code, Version)}{
  	(C, V)
  }  
  \Query {compare(O) : Boolean}{
  	$V \le O.V$
  }  
  \Update{update(v, c)}{
    \textbf{pre} $v > V$\\
    $t :=$ teardown($C$)\\
    $V := v$\\
    $C := c$\\
    init($C, t$)
  }
  \Merge{(N)}{
  	\If{$V \le N.V$}{
        update($N.V, N.C$);
    }
  }
  
  \caption{App CRDT Specification}
  \label{spec:app-crdt}
\end{algorithm}

We will refer to our state-based CRDT as the \emph{App CRDT} from now on.  
Algorithm \ref{spec:app-crdt} shows its specification.
As mentioned, the App CRDT payload is the application code and its version. 

The initial payload consists of version "0.0.0", and \emph{nil}, i.e. we do not provide any application code.

The \emph{update} operation brings up to date the payload with the given version and application code. 
The precondition ensures that the given version is higher than the current one. 

Note that the CRDT can either be updated with a new application code locally, which will then be propagated to other replicas, or it can receive a code update from another replica. 
In both cases, the new version is assumed to be greater than "0.0.0".
Our approach also handles the \textit{teardown} of the old code and the \textit{initialisation} of the new code to enable state migration between the old and new application instances. This is detailed in the following section.

The \emph{getValue} operation just returns the application code. 
Ultimately, the CRDT represents application code, and the version is only metadata used to correctly merge code updates.

The \emph{compare} operation takes two states as arguments and returns true if the first state's version is smaller than or equal to the second state's version. If a compare operation returns false, it implies the first state has a strictly lower version than the second state. 

The \emph{merge} procedure takes a new state N and updates the local state to conform to the new state if it has an appropriate version. 
Concretely, it selects the application state with the newer version.

This means if the replica receives a newer version, the local state will be updated to this newer version. Otherwise, we will not perform an update.

\subsection{Handling State Migration}

Updating an application instance with new code requires careful management of resources and the application state to ensure a correct transition. 
Our approach supports this process by allowing applications to define a \textit{teardown} handler. 
This handler is invoked when a newer version of the application is being deployed.

The \textit{teardown} handler can release resources held by the current version and optionally return data that needs to be migrated to the new code version. 
Developers need to define an \textit{init} handler that takes this optional state and initialises the application.

It is up to the application developers to ensure that the handlers behave consistently across instances. 
While the application state does not always need to be globally consistent, certain scenarios may require maintaining invariants across instances. 
In such cases, developers must design the \textit{teardown} and \textit{init} handlers to uphold these invariants during the migration process.

\subsection{Interoperability Across Versions}
\label{sub:apicompatible}

Although the App CRDT ensures that all application code eventually converges, code updates are not applied instantly. Since the system synchronizes replicas asynchronously, some application instances may be updated sooner. As a result, some instances may run an outdated code version. This is problematic when there are backwards-incompatible software updates. For example, consider that all of the application instances in the system are running version 1.0.0 of an application. The developer now publishes a new version of the application, tagged 2.0.0, with a backwards-incompatible API change. 
While the instances are being updated, some could run the new version 2.0.0 while the rest are still running version 1.0.0. If these application instances interact, this could lead to errors.

\begin{algorithm}
  \SetKw{State}{state:}{}
  \SetKwProg{On}{on}{ :}{}
  \SetAlgoLined
  \State {$Q :=  \emptyset$} \Comment{set of (n, rcv, msg) triplets;} \\ \Comment{rcv: the receiver object , msg: a message}\\
  \State {$C :=  0$} \Comment{C: number}\\
  \State {$V :=  0.0.0$} \Comment{V: semantic version}\\
  \On {receive(r, m)}{
    $v_m := majorVersion(m)$ \Comment{m's major version}
    \uIf{$ v_m = V\vee unversioned(m)$ }{
      deliver(r, m)
    }
    \uElseIf{$v_m < V$}{deliver(r, updateMessage(m))}
    \ElseIf{$v_m > V$}{
        $C := C+1$\\
        $Q := Q \cup \{(C, r, m)\}$
    }
  }

  \On {appVersionUpdated(v)}{
    $V := v$\\
    $q := \{(c,r,m) \; | \; (c, r, m) \in Q\ \cdot majorVersion(m) \le v\}$\\
    \For{(c,r,m) \textbf{in} q \textbf{ordered by} c \textbf{ascending}}{
        receive(r, m)
    }
  }
  \caption{Distributed algorithm at an application instance showing the interplay between the message delivery API (receiver and deliver) and the App CRDT (appVersionUpdated)}
  \label{alg:pureopalgo}
\end{algorithm}

To enable application instances to interact with others  without requiring all updates to be compatible, we introduce a \emph{API-compatible message delivery} layer.

All application-level messages are transmitted with the code version. 
Our system delivers a message to the application instance immediately or temporarily buffers it, depending on the compatibility of the code versions. 

Algorithm \ref{alg:pureopalgo} outlines the interplay between the message delivery layer and the App CRDT at an application instance. 
When an application receives an application message $m$ for an object it hosts $r$, the \emph{receive} event on that instance is triggered. 
The \emph{deliver} function is provided by the platform to invoke the application operation corresponding to $m$ on the receiver object $r$.

The \emph{majorVersion} function extracts the major versions of the message $v_{m}$ so that it can be compared with the local application instance version $V$. 

We distinguish three cases:

\begin{itemize}
    \item $ v_{m} = V $ If the version of the message $v_{m}$ equals the version of the local application code $V$, the sender application instance runs the same version of the code as the receiving one, so the message can be delivered. Similarly, the system delivers messages which do not carry information about a version, i.e. \emph{unversioned message}. An example of an unversioned message is the update message being propagated between App CRDT replicas.
    
    \item $ v_{m} < V $ If the major version of the message is lower than the major version of the local application code, the application instance received a message from another instance running an older code. 
    
    Although the API between these replicas is incompatible, the developer might be able to provide a function that transforms the received message into one compatible with the local API. This transformation can be implemented in the \emph{updateMessage} function provided in the application code. Given a message $m$, \emph{updateMessage} returns a new message compatible with the version of the local application instance.
        
    \item $ v_{m} > V $ If the major version of the message is higher than the major version of the local application code, the system does not deliver the message, as this might cause errors.

    Since we assume eventual-reliable point-to-point channels, the code update to the higher version will eventually be propagated to the local App CRDT.
    While the App CRDT does not receive the update ( and the local application instance does not run the higher code version), the system buffers the application messages received with the $v_{m}$ version in $Q$. More concretely, the messages are first assigned a local message counter and then stored in $Q$ together with the receiver (line 12 in Algorithm \ref{alg:pureopalgo}).
\end{itemize}

    Once the application instance has been updated, the \emph{appVersionUpdated} event is triggered, which triggers \emph{receive} (and eventually \emph{deliver}) for any messages that can be delivered in the order they were stored (lines 17-18). 
    The event will be triggered directly after the initialisation of the new app version (i.e., after \emph{init} in Algorithm \ref{spec:app-crdt}).

Our DSU approach ensures that application instances with incompatible APIs can still coexist on the same network while being updated. The extension also ensures that the system's delivery guarantees remain unaffected. Since the AppCRDT will be updated eventually, the messages stored in $Q$ will also be delivered.

\section{Implementation} \label{sec:implementation}

We implemented our approach in AmbientTalk\cite{ambienttalk}, a programming language designed for distributed systems running on mobile devices. 
AmbientTalk uses an ambient-oriented programming paradigm\cite{AmbientTalk2006}, an extension of object-oriented programming with actor-based constructs to run distributed applications on mobile devices with unstable network connections. 
More concretely, the language comes with built-in peer-to-peer service discovery and a connection-independent failure model.

In the next sections, we will describe the implementation of our approach in AmbientTalk, as well as an additional extension to our implementation that enables code signing for additional security.

\subsection{App-CRDT Implementation}

\begin{lstlisting}[language=AmbientTalk, caption={Simplified Implementation Code for the App-CRDT Prototype Class in AmbientTalk.}, label={list:at-impl}, float]
deftype AppCRDT;

object: {
  import StateBasedCRDTrait;
  ...
  def init(app := nil, requireAPICompatible := false){
    if: (app != nil) then: {
      self.app := app.new(nil, nil);
    };    
    
    if: requireAPICompatible then: {
      self.setVersionMirror();
    };
    
    self.typeName := AppCRDT;
    self.versionMessages := false;
    self.goOnline();
  };
  
  def merge(other) {
    def oVer := other.app.version;
    def sVer := self.app.version; 
    if: (oVer.isNewer(sVer)) then: {
      def result;
      if: (self.app != nil) then: { 
        result := self.app.teardown(); 
      };
      self.app := other.app.new(self.app, result); 
      self.setVersionMirror();
    };
  };
  
  def setVersionMirror() {
    if: (self.mirror != nil) then: {
      def mirror := MessageVersionMirrorModule.enable();
      mirror.version := self.app.version;
      mirror.updateMessage := self.app.&updateMessage;
      self.mirror := mirror;
    };
  };
  ...
};
\end{lstlisting}

Algorithm \ref{list:at-impl} shows the core of the AmbientTalk implementation of our approach. It directly corresponds with the previously described specification, dealing with application versioning and state migration. 

The code defines the AppCRDT prototype class, which can be used by application developers to implement their own App CRDT-based applications. The \texttt{init} method (lines 6-18) takes care of initialising the application, and ensuring that the application CRDT goes online. 
It also takes care of installing a special AmbientTalk \textit{mirror} using \texttt{setVersionMirror} which will ensure that the asynchronous messaging layer of AmbientTalk is extended with the API versioning logic. We will discuss this in more detail in Section \ref{api-versioning-impl}.

The essence of the logic is embedded in the \texttt{merge} method, from lines 20 to 31. This method is invoked whenever code updates arrive and ensures the teardown of the existing state, and initialisation of the new code corresponding to the specification defined in section \ref{sec:specification}.

\subsection{Versioning Co-existence Implementation} \label{api-versioning-impl}

We implemented API-compatible message delivery by installing a new message delivery protocol in AmbientTalk. The essence of the code for this can be seen in Listing \ref{list:at-api-impl}. 
Messages between replicas are tagged with the code version so that the receiver actor know if it either needs to deliver the message or buffer it. If a message is not tagged, the message is always delivered. This is needed to support messages with the \textit{UnVersionedMessage} tag. If the message is tagged, we implemented the three possibilities described in \ref{sub:apicompatible}.

We can achieve this behaviour by levering on AmbientTalk's meta-object protocol through\textit{mirrors}. 
Mirrors are metaobjects that allow developers to inspect and manipulate application objects (called base-level objects) and their runtime behaviour. 
In our case, we use them to inspect messages when they are delivered to an actor's mailbox (lines 13-31). 
With this code actors receiving messages can decide whether to deliver those messages or to buffer them until the correct application version is active. 
Although adding version information to every message sent by an actor may seem like a complex change, the meta-object protocol allows us to implement this behaviour in a single call. 
At line 34 in Listing \ref{list:at-api-impl} we show how the send method is defined in the mirror installed by the extension.

\begin{lstlisting}[language=AmbientTalk, caption={API versioning implementation code in AmbientTalk.}, label=list:at-api-impl, float]
object: {
  deftype UnVersionedMessage <: /.at.lang.types.AsyncMessage;
  deftype VersionManager;
  
  def enable() {..add mirror to actor..};
 
  def createMessageVersionMirror(actor) {
   extend: actor with: {
    def queue := [];
    def version;
    def updateMessage;
    
    deftype Versioned;
    def receive(rec, msg) {
     if: ((version != nil).and: 
      { is: msg taggedAs: Versioned }) 
      then: { 
       self.receiveVersioned(rec, msg); 
      } else: { 
       super^receive(rec, msg); 
      };
    };
 
    def receiveVersioned(rec, msg) {
     def majorIsNewer := 
       version.major >= msg.version.major;
     if: ((version.apiIsCompatibleWith(msg.version)).or: { majorIsNewer }) 
      then: { super^receive(rec, (majorIsNewer.and: { updateMessage != nil })
       .ifTrue: { updateMessage(msg) } ifFalse: { msg });
     } else: { self.queue := self.queue + [[rec, msg]] };
    };
    
    def send(rec, msg) {
     super^send(rec, 
        (is: msg taggedAs: UnVersionedMessage)
        .ifTrue: { msg } 
         ifFalse: { extend: msg with: { |version| } taggedAs: [Versioned] });
    };
    
    def updateVersion(version) {..dequeue operations if version allows..};
   } taggedAs: [VersionManager];
  };
 }
\end{lstlisting}

The send method is a part of the meta-object protocol and is called whenever an actor sends a message to another actor. In our extension, we first check whether the message is tagged with an \textit{UnVersionedMessage} type tag. If this tag is present, we do not add version information to the message. This allows us to make sure some messages are delivered regardless of the implementation version. If the tag is not present, we add a "version" field to the message containing the version of the implementation used by the sender, and we tag the message with the \textit{Versioned} tag. This allows the receiving actor to easily detect whether a message contains version information or not.

To intercept messages when they arrive, we implement the \textit{receive} method on line 14. This method will be called when the actor receives a message. Here we follow the specification in Algorithm \ref{alg:pureopalgo}, and put messages that have a too new version inside of a queue, to be eventually dequeued when the application version is high enough.

\subsection{App-CRDT Usage} \label{sec:ambienttalk-crdt-usage}
Because the CRDT decides when to update the application, it manages the lifecycle of the replicated application by design. To start an application, the app CRDT can be provided with an initial implementation. The CRDT will then start this application and discover other replicas, updating them or receiving updates itself if needed. The implementation provided to the CRDT can be seen as a specification for starting the application. Practically, the implementation needs to be an object with the following fields:

\begin{itemize}
    \item \texttt{version}: This field should contain a semantic version object that represents the version of the implementation. This information is used by the app CRDT to decide how to merge other implementations.
    \item \texttt{init}: This field should contain a method that initializes the application. This function has special semantics in AmbientTalk: whenever we create a new object by calling the \texttt{new} method, AmbientTalk will clone the object and initialize its state by calling the \texttt{init} function. This is also what will happen in our implementation: when the app CRDT wants to start an implementation, it will create a new app object by calling \texttt{new}. Treating the implementation as an object is useful since it allows us to associate the state with the implementation. This can be used later when we want to clean up its resources.
    \item \texttt{teardown}: Whenever the app CRDT receives an implementation with a higher version, it needs to update the application that is running locally with the new implementation. To shut down the old implementation and clean up any resources, the implementation can define a \textit{teardown} function. This is useful when working with external libraries, like GUI frameworks. If an application requires a GUI, the logic for closing the GUI can be provided in the teardown function.
\end{itemize}

Whenever the CRDT wants to create a new application, it will create a new app object by calling the new method on the implementation object. One of our design goals was to provide the ability to migrate the state of an application during an update. This migration logic depends on the application and needs to be provided by the application developer. More specifically, migration logic can be implemented in the initialization method. If the app CRDT is updating an existing application, it will pass the old application object as an argument to the initialization function. This gives the developer access to the old state of the application, allowing it to migrate local state if needed.

Note that this migration logic can be as simple or as complicated as needed. If the application does not have any local state, or if it does not need to be transferred, the initialization function can initialize the new app object as normal. If some state does need to be transferred, the old application object also contains its version. This can be used to customize the migration logic based on the version of the old application.

Listing \ref{list:simple-example} shows an example of how the implementation of an application can be replicated using the CRDT.

\begin{lstlisting}[language=AmbientTalk, caption=A simple application that uses the App CRDT., label=list:simple-example, float]
def SemanticVersion := /.project.lib.semanticversion;  
def AppCRDT := /.project.lib.appcrdt;  
  
def app := isolate: {  
  def version := SemanticVersion.new(1,0,0);
  def state;
  
  def init(previous) {  
    if: (previous != nil) then: {
      self.state := previous.state;
    } else: {
      def random := jlobby.java.util.Random.new();
      self.state := random.nextInt(10);
    };  
  };  
  
  def teardown() { };  
};  
  
AppCRDT.new(app, nil, nil);  
network.online();
\end{lstlisting}

We start by creating our application specification on line 4. Instead of creating an ordinary object, we create an isolated object using the \texttt{isolate:} method. This is related to how AmbientTalk sends objects to other VMs. In our example, we want to send this app object to other replicas. There, the CRDT can use the version field to decide whether to update the implementation or not. If an update is required, it can create an instance of the new application by calling the new method.

However, if we send an ordinary object to another VM, the other replica will receive a \textit{Far Reference} instead. A far reference represents a reference to an object from another actor. This is because each actor has its own, isolated state. Far references do not allow synchronous access to the object they reference. Instead, we can use a far reference to send an asynchronous message to the object. This allows us to easily send objects to other actors or VMs without breaking the actor system. However, this is not what we want in our app CRDT.

When we send the implementation to another VM, we want to receive an ordinary object and not a far reference. The goal of replicating the implementation is that each replica has the implementation locally. We can't use a far reference to instantiate a new application. We can only send messages to it, which will be processed by the actor that contains the object. Fortunately, we can create an isolated object instead. The difference between a regular and an isolated object is that the latter is sent by value. This means that other replicas will receive a copy of the implementation, instead of a reference to the implementation object on another VM.

Our implementation object contains the required \texttt{version}, \texttt{init}, and \texttt{teardown} methods. We also create a \texttt{state} field which the application will use to store some local state. In the initialization function, we can provide our migration logic. If the argument of \texttt{init} is not nil, it will contain the old application object. If we are updating the application, we make sure that the local state is copied correctly. Otherwise, we initialize the state with a random number.

After creating our application specification object, we can start it by creating a new app CRDT and providing it with our implementation object. The CRDT will manage the lifecycle of the application, initializing it on startup and updating it whenever a newer version is received.

This implementation demonstrates that our three design goals have been achieved. First of all, the logic related to communicating with other replicas to receive and send implementation updates is hidden from the user of the CRDT. The specification object of the application only needs to contain the version and the code, while the app CRDT takes care of distributing and updating the implementation. Our second design goal was to provide CRDT users with a way to migrate the local state of an application when it is updated. This is enabled by the initialization method, which can contain complex state migration logic. Our final design goal was a separation of concerns. Although our example is small, the idea is that the core functionality of the application can be located in separate files or modules. For example, we could implement our GUI in a separate object that is created by the initialization method. The GUI implementation does not need to know that its implementation is distributed across multiple replicas. This allows us to easily integrate existing applications with our CRDT.

Although our CRDT can keep an implementation consistent across multiple replicas, it poses some risks when it comes to security. The next section introduces an extension that allows the CRDT to check whether an update was produced by a trusted party.

\subsection{Extension: Code signing} \label{sec:code-signing}
In networks with no authentication or authorization policies, a malicious VM could start distributing updates to other replicas with implementations containing untrusted code. To prevent this from happening, our App-CRDT implementation incorporates code signing functionality.

When the code signing extension is enabled on the CRDT, it will check whether every received update is correctly signed by a trusted party. If the signature is valid, the update is applied (if needed). Otherwise, the message is ignored. We can use asymmetric cryptography with a public-private key pair to generate a signature. We can add a signature by hashing the implementation and encrypting the hash using the private key. Replicas that receive the update can then hash the received implementation, and then decrypt the signature with the public key. If the signature is valid and the implementation has not been tampered with, the decrypted string will be equal to the hash.

This extension allows the replicas to be certain that an implementation was created by a trusted party. A malicious replica could try to change the implementation in a \textit{Man In The Middle} attack. However, this would cause the signature verification to fail on the replica that receives the update. Since the implementation has been modified, its hash will be different, which will cause a mismatch between the hashed implementation and the decrypted signature.

Even though the implementation is distributed in a peer-to-peer fashion, the replicas can verify that an update message is genuine by checking the signature.

\section{Evaluation}
In this section, we evaluate our approach by implementing two applications using our AmbientTalk implementation. Furthermore, we develop a new implementation of our approach in the Lua programming language, demonstrating its portability to other programming environments.

\subsection{Applications in AmbientTalk} \label{sec:apps}

Looking back at listing \ref{list:simple-example}, it shows a very basic example of using the app CRDT. More complex applications can similarly be integrated with the app CRDT. An interesting example of an application whose implementation can be replicated uses CRDTs itself. For example, we could use the app CRDT to replicate the implementation of a state-based counter CRDT. This is discussed in the next section.

\subsubsection{Replicating state-based CRDTs} \label{sec:state-based-crdts}
List \ref{list:at-example-state} shows example code for a basic application that makes use of a state-based counter CRDT. The data type of this CRDT is a counter, but its underlying state uses a mapping from replica IDs to individual counters. Whenever a replica increments the counter, the counter corresponding to its ID is incremented.

Thanks to our design goal of ensuring a separation of concerns, we can implement this counter and a GUI for interacting with the CRDT separately from the implementation distribution logic.

\begin{figure}
    \centering
    \includegraphics[width=0.5\textwidth]{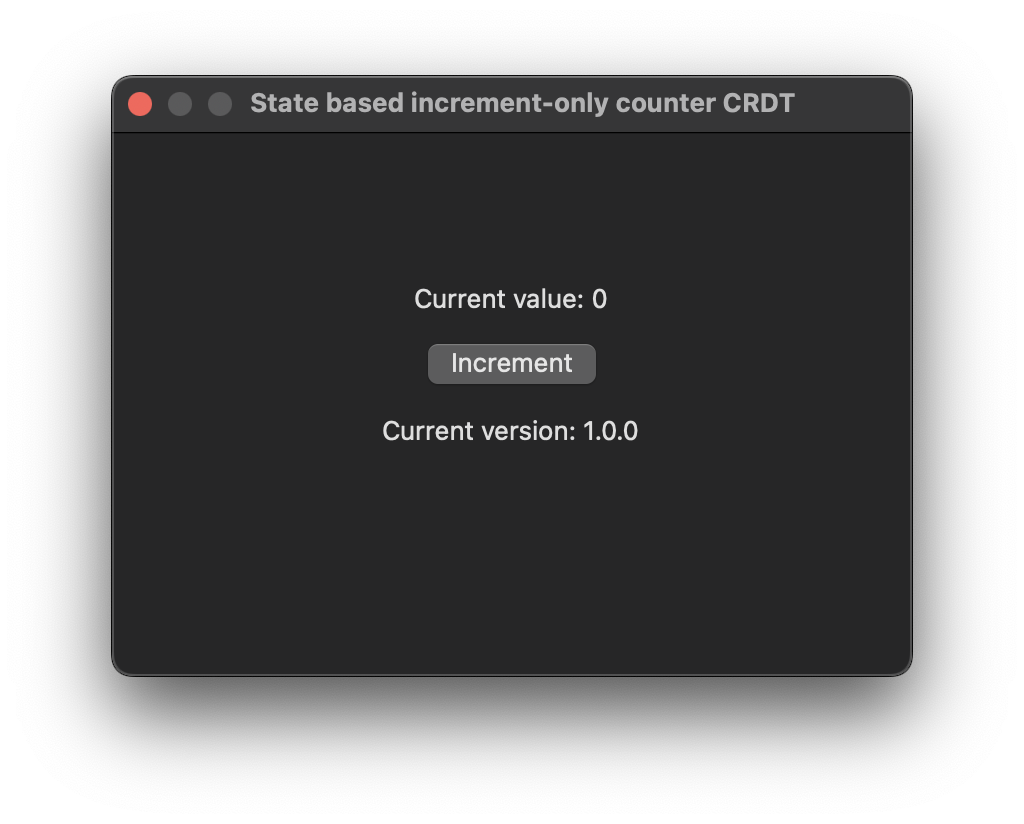}
    \caption{A screenshot of the GUI used to interact with the counter.}
    \label{fig:counter-screenshot}
\end{figure}

\begin{lstlisting}[language=AmbientTalk, caption={Code for state-based app.}, label=list:at-example-state, float]
def app := isolate: {
  def version := SemanticVersion.new(1,0,0);
  
  def counterPrototype := ~.CounterCRDT;
  def guiClass := ~.gui;
  def counter;
  def gui;

  def init(previous, teardownState) {
    self.counter := self.counterPrototype
        .new(getActorId());
    self.gui := self.guiClass
        .new(self.version, self.counter);
    
    counter.wheneverChanged({ |counter| 
      self.gui.updateValue(counter.value());
    });
    
    if: (previous != nil) then: {
      counter.id  := previous.counter.id;
      counter.map := previous.counter.map
    };
    
    self.gui.updateValue(counter.value());
    counter.goOnline()
  };

  def teardown() {
    self.gui.close();
    self.counter.goOffline()
  };
};

AppCRDT.new(app, nil, true);
\end{lstlisting}

Figure \ref{fig:counter-screenshot} shows a screenshot of the GUI that can be used to interact with the CRDT. It shows the current value of the counter, an increment button, and the version of the application that is running.

Because the replicated application now uses a CRDT as well, every VM on which the application is running will now export two CRDTs: one for replicating the implementation of our application, and one for replicating the counter. A difference between both CRDTs is that the app CRDT will remain active as long as the program is running, while the counter CRDT can be shut down and recreated depending on the updates that are received. Because multiple counter CRDTs can be created throughout the lifecycle of the program, we need to ensure that these CRDTs are correctly shut down using the teardown function.

The state of the CRDT consists of a mapping from replica IDs to counters, and the ID of the current replica. Whenever the counter CRDT is updated, this state is copied to the new instance to ensure that the counter remains the same.

In the next section, we discuss how an operation-based CRDT can be replicated using the app CRDT.

\subsubsection{Replicating operation-based CRDTs} \label{sec:operation-based-crdtsr}
An operation-based CRDT updates known replicas by sending the operations performed on the CRDT instead of the entire state, like state-based CRDTs. From an implementation point of view, these approaches are similar. However, some operation-based CRDTs require causal order delivery guarantees.

Causal order delivery can be implemented in AmbientTalk by installing a new actor mirror and changing the behaviour for receiving and sending messages. Similarly to the approach used for implementing API-compatible message delivery, we can include a vector clock in each message. When we receive a message containing a vector clock, we can check whether we can deliver the message based on the internal clock.

Although this implementation of causal order delivery is idiomatic AmbientTalk, it does complicate replicating the implementation itself. This is because the causal order delivery mirror is installed directly on the main actor, so we need to make sure that it is properly uninstalled when the implementation is updated. A second challenge is the fact that this mirror might interfere with our API-compatible message delivery mirror. In this case, the order in which these mirrors are installed matters. For example, if we install the API-compatible message delivery mirror first, followed by the causal order delivery mirror, this will cause the methods of the latter mirror to be called first. This is because when we install a new mirror, we extend the existing one. When a message is received, the send method in the causal order delivery mirror will be called first. However, if the message originated from a replica with an incompatible API version, the causal order delivery mirror may throw errors. For example, if the update changed the API of vector clocks, this could cause issues when the mirror receives a vector clock with a different API.

Therefore, we need to make sure that the API-compatible message delivery mirror is always installed last since this ensures we can buffer messages from incompatible API versions before they are passed on. Making sure the mirror is installed last is not difficult, since we can simply install it after initializing the app. This assumes that mirrors are not installed dynamically in response to certain messages.

To make sure that the mirror is uninstalled correctly, we can store an uninstall function in the mirror itself.

\begin{lstlisting}[language=AmbientTalk, caption=A code snippet that adds an uninstall function to the causal order delivery mirror., label=list:cod-snippet]
def newProtocol := makeCausalOrderingManager(defaultActorMirror, id);
def oldProtocol := defaultActorMirror.becomeMirroredBy: newProtocol;

newProtocol.uninstallCOD := {
  defaultActorMirror.becomeMirroredBy: oldProtocol;
  newProtocol
};
\end{lstlisting}

Listing \ref{list:cod-snippet} shows how an uninstall function for a mirror can be added. We start by creating a new meta-object protocol that enables causal ordering delivery. This protocol extends the default one that is currently installed in the actor. We can use the built-in \texttt{becomeMirroredBy:} method to install a new protocol. Calling this method installs the given protocol in the actor. However, this method also returns the old meta-object protocol.

When we have the old protocol, we can uninstall the new protocol by installing the old protocol again. This approach is sufficient for our use case, where we need to be able to uninstall the causal order delivery mirror when the application is updated. However, this approach also has some flaws: if we extend the actor mirror after creating the uninstall function, uninstalling will remove those mirrors as well. For example, if we call the \texttt{uninstallCOD} function from this example, the API-compatible message delivery mirror will also be uninstalled because it is only installed after the COD mirror. As mentioned, this is not a problem for our use case, but since being able to uninstall specific mirrors is a useful feature, we leave it as future work.

\subsection{LuAT} \label{sec:introduction-to-luat}

We also implemented our CRDT in LuAT\cite{luat}, an ambient-oriented programming framework for Lua. Whereas AmbientTalk can be regarded as a programming language for research purposes, Lua is currently used in real-world systems. Extending our implementation to Lua shows our approach applicability for real-world applications.

As LuAT also implements an ambient-oriented programming model, it shares many features with AmbientTalk. The foundation of AmbientTalk is asynchronous message passing between actors, but Lua does not directly support actors. However, LuAT enables actors to be implemented by providing a set of concurrency-related functions that use Lua coroutines under the hood. As in AmbientTalk, LuAT provides methods to export and discover objects on a local network by connecting to other LuAT programs in a peer-to-peer fashion. We can also create objects in LuAT. Objects can contain state and methods and are transformed into far references when sent to a different actor.

However, we cannot create isolated objects in LuAT. This means that objects sent to other actors are always sent by reference, and cannot be sent by value. A solution to this problem is to send Lua tables instead of LuAT objects. Because a table is seen as a primitive value by the serializer, it is sent by value to other replicas.

Ultimately, LuAT is a framework implemented on top of Lua, while AmbientTalk is a full-fledged programming language. But for our discussion on extending our code replication CRDT to other languages, LuAT is an ideal choice since it also supports the ambient-oriented programming paradigm. In the next section, we will discuss how the app CRDT can be used in LuAT.

\subsubsection{Usage} \label{sec:luat-crdt-usage}
\begin{lstlisting}[language=Lua, caption=A simple LuAT application that uses the App CRDT., label=list:luat-simple-example, float]
local Actor = require "actor"

Actor(function()
  local SemanticVersion = require "lib.semantic_version"
  local App = arequire"project/lib/app_crdt"

  local app = {
    version = SemanticVersion.new(1, 0, 0),

    init = function(self, env, old)
      if old ~= nil then
        self.state = old.state
      else
        self.state = math.random(10)
      end
    end,

    teardown = function(self) end
  }
    
  local appCrdt = App(app, _ENV)
end)
\end{lstlisting}

Listing \ref{list:luat-simple-example} shows a simple program that uses the app CRDT to replicate the implementation of an application. It shares many similarities with the same program implemented in AmbientTalk from listing \ref{list:simple-example}.

In LuAT, we need to create an Actor explicitly by calling \texttt{Actor} with a constructor function. This constructor can return an object containing the actor's behaviour, but since our main actor does not need any behaviour, we do not return anything in the constructor. Instead, we define our application specification in a similar way as in AmbientTalk. We create a table that includes a version of our implementation, an \textit{init} function, and a \textit{teardown} function. This table will be copied by value when sent to other replicas, ensuring that the implementation is correctly replicated.

Based the sample application, we have additionally successfully ported our state-based CRDT counter application from AmbientTalk to LuAT, showing that the proposed code-replicating approach can be generalized to other ambient-oriented programming languages.

\section{Limitations and Future Work} \label{sec:future-work}
This section discusses some limitations and avenues for future work for our DSU approach.

In this work, we assume that each change to an application's codebase is associated with a new version. 
Allowing different code implementations for the same version could lead to nondeterministic behaviour in our current approach, as the outcome would depend on the order in which updates are received. 
One potential solution is to generate a hash of the application code and use it to establish a total order between updates with the same version. 
This would ensure proper convergence since it is all updates are applied in a deterministic order.

Currently, our approach uses a state-based CRDT to replicate the implementation of an application. The downside of using a state-based CRDT is that the entire state is sent to replicas whenever it is updated. This means that the entire implementation must be sent on every update, even if the change itself was only limited to a single line. 
An operation-based approach could be explored to minimize network overhead. 
Instead of sending the entire application code, an operation-based CRDT would only have to send fine-grained code changes. 

However, the operations would have to contain enough information for the replicas to update the local application instance correctly, and the programming language should have powerful meta-programming support that allows us to record code changes, and apply the changes on another application instance.

Our approach also does not allow the implementation of the App CRDT framework itself to be updated dynamically. The DSU literature has identified that updating the DSU itself is difficult\cite{Polito15}. Currently, should changes be required to the App CRDT, all replicas must be manually restarted with the updated App CRDT implementation. This could be solved by allowing the App CRDT to receive updates. However, careful consideration is needed to allow different versions of the App CRDT to co-exist and avoid meta-circularity problems.

In the current implementation of API-compatible message sending, messages that cannot be delivered due to version mismatches are buffered until the application is updated. However, some messages may remain backwards-compatible even after a major version upgrade, allowing them to be delivered immediately while others are buffered. 
To enhance flexibility, developers could specify the desired behaviour for individual message types using a message type tag (such as in AmbientTalk). At present, messages can be marked as \textit{unversioned}, ensuring they are always delivered regardless of version changes. 
Expanding this approach to allow messages to be tagged with compatible version ranges could provide more granular control and enable improved functionality in handling version mismatches. 
Another approach would be to use static analysis tools to determine whether the updated version has backwards-incompatible API changes \cite{apidiff}.

\section{Conclusion} \label{sec:conclusion}
In this paper, we introduced a novel approach to dynamic software updates (DSU) in distributed systems using Conflict-free Replicated Data Types (CRDTs)
We introduce the \emph{App CRDT} for replicating code in a distributed system while tracking versions. 
Our approach supports state migration, allowing developers to define how application instances can transition correctly between versions while maintaining essential invariants. 
Additionally, our approach does not require all updates to be compatible, but it enables the interoperability of application instances running different code versions through an \emph{API-compatible message delivery}.

We demonstrated the practicality of this approach through an implementation in the AmbientTalk programming language and validated its portability by creating a secondary implementation in LuAT for Lua.  
These implementations highlight the flexibility and generalizability of the App CRDT across programming languages and runtime environments. 
We validate these implementations through the implementation of several sample examples, hosting for example state-based and operation-based CRDTs. 

Future work includes exploring operation-based CRDTs for code replication to reduce network overhead, addressing dynamic updates to the App CRDT framework itself, and improving message compatibility through finer-grained control over versioned communication.

\bibliographystyle{ACM-Reference-Format}
\bibliography{references}

\end{document}